\documentclass[%
 reprint,
 superscriptaddress,
 amsmath,amssymb,
 aps,
prb,
]{revtex4-1}

\usepackage{graphicx}
\usepackage{dcolumn}
\usepackage{xcolor}
\usepackage{bm}
\usepackage{esint} 

\setcitestyle{numbers,square}

\begin{document}

\preprint{XXXX}

\title{
First principles calculations of
steady-state voltage-controlled magnetism:
application to x-ray absorption spectroscopy experiments
}

\author{Alberto Marmodoro}
\affiliation{
Ludwig-Maximilians-University Munich,
Department of Chemistry,
Butenandtstrasse 11,
D-81377 Munich,
Germany
}%
\affiliation{FZU -- Institute of Physics of the Czech Academy of Sciences, Cukrovarnicka 10, CZ-162 53 Prague, Czech Republic}
\author{Sebastian Wimmer}
\affiliation{
Ludwig-Maximilians-University Munich,
Department of Chemistry,
Butenandtstrasse 11,
D-81377 Munich,
Germany
}%
\author{Ond\v{r}ej \v{S}ipr}
\affiliation{FZU -- Institute of Physics of the Czech Academy of Sciences, Cukrovarnicka 10, CZ-162 53 Prague, Czech Republic}
\author{Masako Ogura}
\author{Hubert Ebert}
\affiliation{
Ludwig-Maximilians-University Munich,
Department of Chemistry,
Butenandtstrasse 11,
D-81377 Munich,
Germany
}

\date{\today}%

\begin{abstract}
Recent x-ray absorption experiments have demonstrated 
the possibility to accurately monitor the magnetism
of metallic hetero-structures controlled via a 
time-independent perturbation caused for example by a static electric field.
Using a first-principles, non-equilibrium Green function scheme, 
we show how the measured dichroic signal for
the corresponding  steady-state situation can be related 
to the underlying electronic structure and its response to
the external stimulus. The suggested approach works 
 from the infinitesimal limit of linear response
to the regime of strong electric field effects, 
which is realized in present experimental high sensitivity investigations.

\end{abstract}

\pacs{Valid PACS appear here}

\maketitle

The creation, control and detection of spin polarized currents 
lie at the heart of spintronics.
Accordingly,  a number of magneto-optical
 experiments have been proposed over the years
to explore related magnetic phenomena and corresponding materials. 
Obviously,
X-ray-based measurement techniques seem to be especially suited
for this purpose because they provide a high signal-to-noise ratio, 
precise targeting of individual chemical species
and tunable sensitivity to either surface or bulk-like properties of a sample,
depending on photon energy \cite{VanderLaan2017,Bonetti2017}.

In particular the
 time-dependence of  magnetic properties has been studied 
 over a broad range of time scales by means of  magneto-optical techniques, addressing for example 
ultra-fast demagnetization  \cite{Beaurepaire1996,Ghiringhelli2001,Rhie2003,Battiato2010,Koopmans2010,Mann2012,Hofherr2017}
or optical manipulation of the magnetic order \cite{Kirilyuk2010,Elliott2016,Ishii2018,Dewhurst2018}.
In general, the most comprehensive information
can be acquired by corresponding experiments using pump-probe techniques.
Early experiments in the field, such as time-dependent 
MOKE (magneto-optical Kerr effect) or 
 XMCD (X-ray magnetic circular dichroism)
experiments
\cite{Beaurepaire1996,Ghiringhelli2001} have been performed 
by controlling the investigated system  via an external magnetic field
and using different delay times for the read-out via x-ray pulses.

Another type of interesting 
experiments exploits the XMCD to study the impact of an
external static electric field on the magnetization.
For the resulting  out-of-equilibrium  situation
one may have pure charge rearrangements 
with  a  continuous flow of  electric charge   prevented by  applying the electric
field across  an interface to vacuum or  an insulating layer,
i.e.\ having a capacitor like experimental setup \cite{Miwa2017,Yamada2018}.
For the combination of conducting subsystems,
on the other hand, a steady-state  out-of-equilibrium situation
will be created  with a constant electric current flowing
\cite{Kukreja2015,Stamm2019}.
In this case one may focus on the
electric field induced change of the magnetization 
longitudinal \cite{Kukreja2015} or transverse  \cite{Stamm2019} with respect to the  electric field. 
The electric field-induced  electric current will in general be accompanied by a spin current that might be used for example for switching the magnetization
via the spin transfer torque (STT),
spin-orbit torque (SOT)
or the spin Hall effect (SHE) \cite{TZ19}. Accordingly, the 
question concerning the connection between the observed XMCD
and the induced spin current arises in a natural way.

In the following,
a theoretical description 
of the electric field induced XMCD for the case
of a conducting system is given, with a focus on the 
longitudinal setup investigated by Kukreja \textit{et al.} \cite{Kukreja2015}.
These authors investigated the
bilayer system Co/Cu by XMCD measurements at the Cu L$_3$-edge.
It was found that a voltage applied across the
layer system changes the XMCD spectra primarily in the
vicinity of the Fermi level. By switching the sign of
the voltage it was possible to separate in a
reliable way the electric field induced
contribution from that due to the
so-called proximity effect (see below).
This allowed in particular to demonstrate
the linear dependence of the induced  changes
for the XMCD spectra on the applied
voltage.
Here, we present
an ab-initio description of the observed
phenomenon, that accounts in particular  for the
out-of-equilibrium situation
when a finite voltage is applied to a conducting system.
To deal with the XMCD in this case, an appropriate
expression for the X-ray absorption
coefficient together with a
corresponding extension of the
XMCD sum rules \cite{Wienke1991,Schutz1993,Thole1992,Carra1993}
are suggested. 
This allows to relate the XMCD spectra of an atom to its spin and
orbital moments under the new out-of-equilibrium conditions.

\medskip

\begin{figure}[htb]
\centering
\includegraphics[width=8cm]{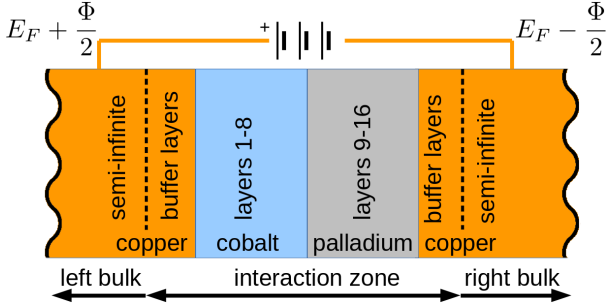}
\caption{\label{fig: device}Sketch of the layered system considered: a Co/Pd layer system is connected to the left
and to the right to
Cu leads. The Co and Pd layers, together with few neighboring
Cu layers, form the central so-called interaction zone.
}
\end{figure}

Considering the electronic structure of a system 
out-of-equilibrium,  the 
occupied and unoccupied states can be represented in
an appropriate way in terms of the lesser and greater Green
functions, respectively \cite{Stefanucci2013}.
Here we deal with the steady state situation 
encountered for a 
layered system exposed to a constant
 electric field across the layers,
or equivalently a layered system connected to left and
right leads with a corresponding
voltage drop $\Phi$ in between.
In this case, it is most convenient to 
consider the lesser and greater Green functions, 
$G^<_{\Phi}(E)$ and $G^>_{\Phi}(E)$, respectively, as functions of
the energy $E$ and dependent on the applied voltage  $\Phi$.  
For the  setup of a layered system as sketched schematically
in Fig.\ \ref{fig: device} it was shown by several authors
\cite{Henk2006,Heiliger2008a,Achilles2013}
that the Green functions 
$G^{\lessgtr}_{\Phi}(E)$  can be calculated by means of a 
 Dyson-like equation, 
 that relates the various spatial regions
 and Green functions,
through an appropriately defined  self-energy
${\Sigma}^{\lessgtr}_{\Phi}(E)$. 
%
%
%
This way one gets  $ G^{\lessgtr}_{\Phi}(E)$   by
the expression \cite{Wang2009a,Kubis2011,Pourfath2014}
\begin{equation}
{G}^{\lessgtr}_{\Phi}(E)
=
{G}^{r}_{\Phi}(E) \,
{\Sigma}^{\lessgtr}_{\Phi}(E)\,
{G}^{a}_{\Phi}(E)
\end{equation}
 in terms
of the retarded and advanced Green functions,
$G^{r}_{\Phi}(E)$  and $G^{a}_{\Phi}(E)$, respectively,
that in turn can be calculated using standard techniques \cite{EKM11}.
In this context, the finite voltage 
$\Phi$ determines the finite and fixed difference
in the chemical potentials of the left and right leads,
$\mu_L=E_F + \Phi/2$ and $\mu_R=E_F - \Phi/2$, respectively,
where $E_F$ is the Fermi level of the unperturbed reference
system with $\Phi=0$~V.
To deal with the gradual voltage drop across 
the region between the leads, 
denoted interaction zone in Fig.\ \ref{fig: device},
 a corresponding model suggested in the literature \cite{Taylor2001,Brandbyge2002}
 was adopted.
(See Appendix~\ref{sec-voltage} for more details).
 
The scheme 
to calculate the lesser and greater Green functions, 
briefly sketched  above, was implemented 
within the framework of local spin density approximation
making use of the Korringa Kohn Rostoker (KKR)
Green function technique as suggested by Ogura and Akai
\cite{Ogura2016}.
As the study aimed among others
 at the calculation of 
spin-orbit induced XMCD spectra,
a fully relativistic formulation
 of the scheme was used on the basis of the four-component Dirac equation, i.e.\
 all  Green functions are $4 \times 4$-matrix functions
 \cite{EBKM16}.

  With the lesser and greater Green functions
$G^{\lessgtr}_{\Phi}(E)$ available, 
many electronic properties of interest can be 
calculated straightforwardly. For example the 
density of states  (DOS)
 related to the occupied $(<)$ and unoccupied  $(>)$ states is given 
 by $n_{\Phi}^{{\lessgtr}(d)}(E) = -1/\pi \,\mbox{Trace} \, G_{\Phi}^{{\lessgtr}(d)}(E)$, where the angular-momentum representation used for the 
Green function 
 \cite{EBKM16} allows in a simple way to project out the contribution
 due to d-electrons as indicated by the optional
 index $d$. In a similar manner one has for the 
 spin magnetic moment 
 $m_{\Phi}^{(d)}$ associated with the occupied states
\begin{equation}
\label{eq: Dirac spin-projected IDOS difference}
m_{\Phi}^{(d)}
=
-\frac{1}{\pi}
\int_{-\infty}^{E_F + |\Phi|/2} 
dE\,
\mbox{Trace} \,
\beta \
\sigma_z \,  G_{\Phi}^{<(d)}(E) \;,
\end{equation}
where  $\beta$ 
  is one of the standard Dirac  matrices,  $\sigma_z$ a $4 \times 4$-Pauli matrix \cite{Ros61}
  and the integration regime has been  restricted 
  according to the maximum of 
$\mu_L$ and $\mu_R$
(See Appendix~\ref{sec-practical} for more details.)

For the steady-state situation considered here,
an appropriate 
expression for the X-ray absorption coefficient 
$\mu^{\vec q,\lambda}_{\Phi}(\omega)$ can
straightforwardly be derived 
by replacing   in the standard expression 
for an unperturbed ground state \cite{Ebert1996}
the product  of the retarded Green function and the Fermi distribution
at given temperature, i.e.\  $G^r(E) \, [1-f_T(E)]$, by the greater  
Green function  $G^>_{\Phi}(E)$ representing the unoccupied 
final states.
This leads to:
\begin{equation}
\label{eq:abs-coef}
\mu^{\vec q,\lambda}_{\Phi}(\omega)
\propto
\sum_{i \in occ}
\langle \Psi_i | X_{\vec q,\lambda}^{\dagger} \,
G^{>}_{\Phi}(E_i + \omega)  \,
X_{\vec q,\lambda} | \Psi_i \rangle
\;,
\end{equation}
where   $\Psi_i$ represents an occupied initial core state $i$
while
$X_{\vec q,\lambda}$ stands for 
the interaction of the electrons with the 
photons with wave-vector $\vec q$, energy $ 
\omega$ and polarization $\lambda$ \cite{Ebert1996}.
Dealing with $X_{\vec q,\lambda}$ we make use of the 
dipole approximation. Accordingly, the index 
$\vec q$ can be omitted in the following.
Finally, for the case $\Phi=0$~V the rather general expression 
for $\mu^{\lambda}_{\Phi}(\omega)$ in Eq.\ (\ref{eq:abs-coef}) is of course 
fully equivalent to the standard one in terms of $G^r(E)$  \cite{Ebert1996}.
This is demonstrated by a numerical example in Appendix~\ref{sec-abs}.

An extremely  attractive tool to interpret
 experimental as well as theoretical XMCD data,
i.e.\  the difference in absorption 
of left or right circularly polarized x-rays,
is provided by the XMCD 
sum rules
\cite{Wienke1991,Schutz1993,Thole1992,Carra1993}.
%
Focusing on  $L_{2,3}$-spectra, 
the spin-related sum rule  is given by
\begin{widetext}
\begin{equation}
\label{eq:sum-rule}
\frac{ m^d + 7 \times T^d_z }{ N^d_h }
=
\frac{
  \int_{E_{\rm min}}^{E_{\rm cutoff}} 
  \left[
    \mu^{\lambda=+}_{L_3}(\omega)
    -
    \mu^{\lambda=-}_{L_3}(\omega)
  \right] d\omega
- 2 \times 
    \int_{E_{\rm min}}^{E_{\rm cutoff}} 
    \left[
      \mu^{\lambda=+}_{L_2}(\omega)
      -
      \mu^{\lambda=-}_{L_2}(\omega)
    \right] d\omega 
}{
\int_{E_{\rm min}}^{E_{\rm cutoff}} 
\left[
\bar \mu^{}_{L_3}(\omega)
+
\bar \mu^{}_{L_2}(\omega)
\right] d\omega
},
\end{equation}
\end{widetext}
where $m^d$ is the spin magnetic moment
associated with d-states,
$T_z^d$ denotes the magnetic dipole or asphericity term,
$N^d_h = 10 - \int_{-\infty}^{E_F} n^d(E) \, dE
$ 
is the number of $d$-like holes 
derived from the $d$-like DOS $ n^d(E)$
above the Fermi level $E_F$
and 
$\bar \mu^{}_{L_{2(3)}}(\omega)$ stands for the
polarization averaged 
absorption coefficient for the $L_2$ ($L_3$) edge.
All energy integrals in Eq.\ (\ref{eq:sum-rule})
run up to  the cutoff energy $E_{\rm cutoff}$
for which the integrated d-like DOS 
covers  10 d-states; i.e.\
$
\int_{-\infty}^{ E_{\rm cutoff} }
n^d(E) \, dE = 10$.

Equation \eqref{eq:sum-rule} was initially derived
for the situation that no  external voltage is applied
to the system.
Nevertheless, it can be straightforwardly  adapted to the 
situation  of a finite applied voltage  $\Phi$
by calculating the absorption coefficients 
$\mu^{\lambda}_{\Phi}(\omega)$
by means of Eq.\ (\ref{eq:abs-coef})
and using the greater Green function  $G^>_{\Phi}(E)$.
The quantities $m^d$, $T_z^d$ and $N^d_h$,
on the other hand,
have to be calculated accordingly using  
the lesser Green function  $G^<_{\Phi}(E)$.
Finding the number of $d$ holes $N_h^d$ 
requires in particular
an integration of the  DOS $n^{<d}_{\Phi}(E)$ 
related to $G^{<d}_{\Phi}(E)$
up to the energy where it goes to zero.
This upper  threshold energy  lies typically above the original
Fermi energy  $E_F$  for the case  $\Phi=0$~V.
On the other hand, 
the lower 
 threshold energy from which on unoccupied states may 
contribute to the X-ray absorption   via  $G^>_{\Phi}(E)$ 
is given by 
$E_{\rm min} \leq E_F - \left|\frac{\Phi}{2}\right|$.

\bigskip

To avoid numerical problems and to ensure 
pronounced field-induced changes of the magnetic properties
the   bilayer system  Co/Pd  has  been considered in this
work, as Pd has a much higher spin polarizability than Cu.
Otherwise, the set-up sketched in Fig.\ \ref{fig: device}
followed essentially the work of 
Kukreja \textit{et al.} \cite{Kukreja2015} on Co/Cu; i.e.\
we investigated 
 a fcc (001) textured Co/Pd bilayer system consisting of
8 layers of Co with their magnetization oriented perpendicular 
to the layers (out-of plane) and 8 layers of Pd.
On both sides 4 buffer layers of Cu have been added to allow for a
smooth connection to the fixed Cu leads
with their respective chemical potential shifted by the
voltage drop $\Phi$.

\begin{figure}[htb]
\centering
\begin{tabular}{c}
\includegraphics[width=7.5cm]{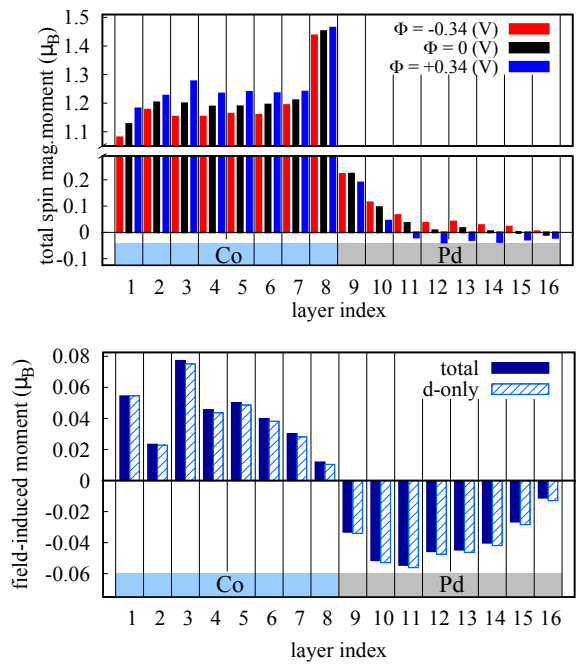}
\end{tabular}
\caption{\label{fig: SMT vs. IMQ}
  Top: layer ($n$) resolved profile of the spin magnetic moment 
  $m^{n}_{\Phi}$ (in $\mu_B$) for the
  investigated Co/Pd bilayer system
for three applied voltages: 
$\Phi$ = $-0.34$  (red bars), $0$  (black bars) and $+0.34$ V (blue bars).
Bottom: electric field induced contribution to the profile 
$\Delta m^{n}_{\Phi} =  m^{n}_{\Phi} -  m^{n}_{0}  $  
(full bars)
for the voltage drop $\Phi=+0.34$~V with 
$\Delta m^{d\,n}_{\Phi}$ its part due to the d-electrons (shaded bars).  
}
\end{figure}

The top panel of 
Fig.\ \ref{fig: SMT vs. IMQ} shows the resulting
profile of the  spin magnetic  moment $m^{n}_{\Phi}$ for the Co/Pd bilayer
for  $\Phi=0$ and $\pm 0.34$~V, where $n$ is the layer index.
%
%
For $\Phi=0$~V one finds an  enhancement for the Co moments 
at the Co/Pd interface \cite{JRKH15},
while for the Pd layers there is an appreciable induced moment
that decays rapidly with the distance from the interface (proximity effect).
These well known features of the magnetization
profile at the Co/Pd interface are
obviously essentially unchanged when a finite voltage  $\Phi$ is applied,
that modifies the magnetic moments depending on the sign of $\Phi$.
The corresponding  electric field induced contribution to the
magnetization profile 
$\Delta m^{n}_{\Phi} =  m^{n}_{\Phi} -  m^{n}_{0}  $
is shown for  $\Phi=+0.34$~V
in the bottom panel of Fig.\ \ref{fig: SMT vs. IMQ}.
As one can see, the changes of the  Co and Pd moments are
of the same order of magnitude but interestingly of
different sign. In addition one notes that 
$\Delta m^{n}_{\Phi} $ has nearly exclusively to be ascribed to the
part of the d-electrons ($\Delta m^{d\,n}_{\Phi}$).
The layer resolved induced magnetic moments, and with these also
the averaged induced moments of Co and Pd, scale fairly
well linearly  with the applied voltage $\Phi$
(see below).

\begin{figure}[htb]
\centering
\includegraphics[width=9cm]{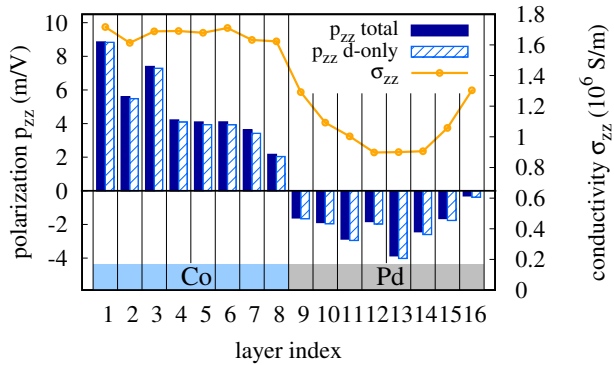}
\caption{\label{Fig: linear response plot}
  Layer $n$ resolved Edelstein coefficient 
  $p_{\rm zz}^{n}$ (full bars) together with  its
part  $p_{\rm zz}^{d,n}$ 
due to the d-electrons (shaded bars). In addition the 
layer resolved charge conductivity  $\sigma_{\rm zz}^{n}$ (orange circles)
is shown. }
\end{figure}

These observed properties of the  induced magnetic moments
suggest an alternative
description of the phenomenon of an
electric field induced magnetization
by means of Kubo's linear response theory. 
The corresponding
Kubo-Bastin like expression for the non-local
response  coefficient $p^{nn'}_{\rm zz}$
that gives for layer $n$ the induced magnetization 
along the z-axis due to an electric field in layer $n'$
along  the  same direction
can be found in Appendix~\ref{sec-edel}.
Corresponding numerical results for the local
 layer dependent
polarization coefficient $p^n_{\rm zz}$
defined by the sum $\sum_{n'}  p^{nn'}_{\rm zz}$
 are given in
 Fig.\ \ref{Fig: linear response plot}
 for the investigated
Co/Pd bilayer.
%
%
Obviously, the profile of $p^n_{\rm zz}$ is in full accordance
with the profile of the induced magnetic moment $\Delta m^{n}_{\Phi}$
shown in Fig.\ \ref{fig: SMT vs. IMQ}.
In particular, the different sign
for the induced moment for Co and Pd is fully confirmed by the linear
response calculations. 
In addition, we find also that  $p^n_{\rm zz}$ 
is by far dominated by its part stemming from the d-electrons  
 ($p^{d\,n}_{\rm zz}$).
Here it should be stressed that a perfect
one-to-one correspondence of
$\Delta m^{(d)\,n}_{\Phi}$ in  Fig.\ \ref{fig: SMT vs. IMQ}
and
$p^{(d)\,n}_{\rm zz}$ in Fig.\ \ref{Fig: linear response plot},
respectively, 
cannot be expected as the evaluation
of the induced magnetic moment
requires the  non-local
response  coefficient $p^{nn'}_{\rm zz}$ together
with the local electric field  $E^{n'}_{\rm z}$
for the various layers. However, the later one will depend in
a non-trivial way on the layer index $n'$ as it is suggested
by the local electrical conductivity  $\sigma_{\rm zz}^{n}$
that varies in an appreciable way from layer to layer
for the Pd subsystem
as can be seen in Fig.\ \ref{Fig: linear response plot}
(orange circles).
Nevertheless, 
the comparison of the profiles of $\Delta m^{(d)\,n}_{\Phi}$ 
and $p^{(d)\,n}_{\rm zz}$ in  Figs.\ \ref{fig: SMT vs. IMQ}
and  \ref{Fig: linear response plot}, respectively,
clearly shows that the electric field induced magnetization
is first of all  a manifestation of the
Edelstein effect \cite{Edelstein1990,Hellman2017}.
Concerning this, it should be stressed that
the lack of inversion symmetry
as the central precondition
for the occurrence of the Edelstein effect   \cite{Edelstein1990}.
This is of course given for the Co/Pd interface region.

\bigskip

The proximity effect mentioned above,
i.e.\ the occurrence of an induced spin magnetic magnetization in
an non-magnetic metal at the interface
to a ferromagnetic metal,
could be demonstrated in the past for several systems
by means of XMCD measurements on the non-magnetic metal.
A prominent examples for this is Pt in  
 Co/Pt  \cite{RSF+91}
but also Cu in  Co/Cu \cite{Samant1994}
with the high and small, respectively, spin susceptibility
of the non-magnetic metal reflected
by the correspondingly high and small  induced spin magnetic moments
that differ by one to two orders of magnitude (see also Fig.\ \ref{fig: SMT vs. IMQ}).

\begin{figure}[htb]
\includegraphics[width=7.5cm]{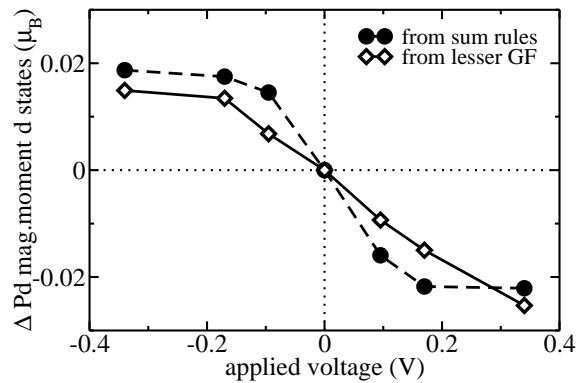}
\caption{\label{fig:Pd-SMT-DOS-XMCD}
  Electric field induced
spin magnetic moment of the  $d$ states of Pd,
as a function of the applied voltage $\Phi$,
averaged over the Pd layers: $\langle \Delta m^{d}_{\Phi} \rangle$
calculated directly  from the lesser Green function
$G^<_{\Phi}(E)$ via Eq.\ \eqref{eq: Dirac spin-projected IDOS difference} (open diamonds)
and
$\langle \Delta m^{{\rm XMCD}\,d}_{\Phi} \rangle$ deduced from
the XMCD spectra at the L$_{2,3}$-edges via the modified
sum rules (see Eq.\ \eqref{eq:sum-rule}) (full circles).
}
\end{figure}

In an earlier study we predicted that the magnetization induced by an
external magnetic field should give rise to a corresponding XMCD
signal that  scales in a one-to-one manner with the spin and orbital
susceptibilities \cite{EM03}.
This could indeed be demonstrated experimentally for the non-magnetic
metals Pt and Pd  \cite{RWJ+06} but also for  Au  \cite{SKM+12}.
In a completely analogous way, one can expect that the magnetization
induced by an external electric field via the Edelstein effect
leads also to the occurrence of an XMCD
in case of an otherwise non-magnetic system.
For an element in a system with a spontaneous or induced magnetization,
on the other hand,
this mechanism will alter the already present XMCD spectrum accordingly.
This was indeed found when calculating the Pd L$_{2,3}$-spectra for the
investigated Co/Pd bilayer system when a finite voltage drop $\Phi$ is applied
(see Appendix~\ref{sec-sumrule}). 
To deduce the relationship of the additional XMCD signal
and spin magnetic moments
induced by the electric field,
the L$_{2,3}$ XMCD spectra for the individual Pd layers
have been calculated in a first step for the voltages $\pm \Phi$.
Taking for each layer the difference  leads accordingly to the field induced contribution of
the XMCD spectra, that have been analyzed by making use
of the modified version of the spin sum rule given in Eq.\ (\ref{eq:sum-rule}).
This leads finally to  the field induced spin magnetic moments
$\Delta m^{{\rm XMCD}\,d\,n}_{\Phi} $ of the d-electrons
as deduced from the XMCD spectra.
As a corresponding experiment does not allow to distinguish the XMCD spectra
of the  individual Pd layers $n$, the average over all these have been taken.
This average value $\langle \Delta m^{{\rm XMCD}\,d}_{\Phi} \rangle$
is given in Fig.\ \ref{fig:Pd-SMT-DOS-XMCD} as a function
of the applied voltage $\Phi $ with the corresponding
averaged spin magnetic moment  $\langle \Delta m^{d}_{\Phi} \rangle$
calculated directly from the lesser Green function
$G^< (E)$  via Eq.\ \eqref{eq: Dirac spin-projected IDOS difference}.
%
%
%
Keeping in mind the various approximations used to derive the
XMCD sum rules   \cite{Ebert1996} the agreement between the directly and
XMCD-derived moments is rather satisfying. This finding is very similar
to the results of studies on systems with the magnetization occurring
spontaneously or induced by the proximity effect  \cite{GETD94}.

The close connection between the spin magnetic moment
$\langle \Delta m^{d}_{\Phi} \rangle$
calculated directly for the Pd layers and
$\langle \Delta m^{{\rm XMCD}\,d}_{\Phi} \rangle$ deduced from
the spectroscopic data clearly shows that the
electric field induced XMCD indeed reflects in a
one-to-one manner the induced spin magnetic moments
for a steady-state situation.
Accordingly, we conclude that the experimental findings of
Kukreja \textit{et al.} \cite{Kukreja2015} for Co/Cu
have to be interpreted as well this way; i.e.\ that the
observed additional XMCD signal due to the applied voltage
has to be seen as a rather direct measure for the electric field  induced spin magnetic moment
and as a manifestation of the Edelstein effect.
The fact that the calculated as well as the observed field induced changes
of the XMCD spectra occur primarily in the vicinity of the Fermi level
supports this conclusion and interpretation of the experimental findings.
Nevertheless, it should be mentioned that the observed XMCD spectra may have
additional contributions due to spin accumulation caused by the
specific features of the experimental set-up not accounted for within the
present theoretical study.

\bigskip

In summary, a numerical study on the electric field induced changes 
in the electronic, magnetic and spectroscopic properties of the 
ferromagnet/non-magnet bilayer system Co/Pd has been presented. 
The field induced magnetic moments were found to scale essentially
 linearly with the applied voltage. This finding as well as 
additional linear response calculations allow to ascribe the 
induced magnetic moments to the Edelstein effect.
 To make  contact with recent XMCD investigations on Co/Cu
an appropriate expression for the calculation of the X-ray absorption
coefficient for a steady-state
out-of-equilibrium situation was suggested.
Together with a corresponding extension of the
XMCD sum rules the one-to-one
corresponding relation between
the electric field induced contributions
 to the spin magnetic moment and XMCD spectra could be demonstrated.
From this it could be concluded that
the results of a recent experimental XMCD study on the bilayer system
Co/Cu first of all
reflect the  field induced magnetic moment in the vicinity of the Co/Cu
interface.

\bigskip

We gratefully acknowledge insightful discussions with Prof.\
Hisazumi Akai, Institute for Solid State Physics (ISSP), Tokyo.
This work was supported by the Deutsche Forschungsgemeinschaft
grant: DFG EB 154/35.

\appendix

\section{Accounting for finite voltage effects} 

\label{sec-voltage}

In order to include effects from a non-zero applied voltage,
the initially equal chemical potential 
of the left and right leads of the considered trilayer system
is rigidly shifted by $\pm \frac{\Phi}{2}$ with respect to 
the Fermi level $E_{F}$ coming from 
 selfconsistent calculations 
 performed without any external perturbation, 
i.e., in terms of the retarded Green function (GF) alone.
This translates into 
 a constant shift 
for the 
 potentials which describe atoms 
of the left or right semi-infinite leads.

Different levels of self-consistency can be deployed
to study the effects of such boundary conditions 
on the quantities of interest within the interaction zone,
i.e., the middle part of the considered trilayer system.
In this work we rely on a ``ohmic conductor'' assumption,
that makes use of Ohm's law $V = R \times I$ 
to simply extend the same manipulation of the potentials
also to the interaction zone. 
The layer-dependent voltage offset
is applied gradually
in a piecewise constant way, so that the whole interval
  $\left[ -\frac{\Phi}{2}, +\frac{\Phi}{2} \right]$
  is covered accross the interaction zone. 
 The difference between the offsets for adjacent layers is, in
  principle, proportional 
to the local resistivity, which undergoes only 
relatively minor variations across the interaction zone.

\section{Practical calculation of observables}

\label{sec-practical}

When evaluating the induced spin moment via Eq.~(2) in the main text, 
the lesser GF goes by definition to zero for unoccupied states.
The choice of the upper integration limit is hence not crucial, 
provided that it samples a large enough energy range
up to which the applied voltage 
can shift spectral weight, with respect to the original Fermi level
for $\Phi=0$~V.
This interval has a width 
roughly proportional to the magnitude of the external perturbation:
$\left[ -\frac{\Phi}{2}, +\frac{\Phi}{2} \right]$,
with a cutoff that becomes smoother with larger electronic temperature (if applied)
in the Fermi-Dirac distribution
that enforces the different definition of
lesser and greater GFs in the steady-state  non-equilibrium Green
function (NEGF) scheme of
Refs.~\cite{Achilles2013,Ogura2016}. 

 Obtaining the observables within the 
NEGF formalism requires that the energy integrations [as in
  Eq.~(3) of the main paper] are performed on the real axis.  One
cannot make use of the complex contour integration as it is common
when dealing with $G^{r}$ or $G^{a}$.
This makes both the Brillouin zone and energy integrations 
numerically quite demanding.
 One has to address numerical problems arising from the need to 
evaluate the relatively small differences between the
majority and minority integrated density of states (IDOS).

This difficulty can be addressed by exploiting  the fact
that changes in the electronic structure due to the NEGF self-energy
are typically significant only close to the Fermi level.
Following proposals in the literature \cite{Achilles2013},
we split the energy integration for the spin projected IDOS
over two sub-intervals,
 one of them covering most of the energy range up to a small
  value below $E_{F}$ and the other one covering the rest. 
Within the  first interval (lying lower in energy) 
the lesser GF  can be safely replaced
by the retarded GF, computed for the same 
local potentials, as discussed in Appendix~\ref{sec-voltage} above. 
This allows to resort to well-known 
Gaussian-Legendre quadrature in the upper complex semiplane
and invest more computational resources 
for the adjacent energy window at higher energy, where we sample the lesser GF
along a contour parallel and close to the real energy axis.
Although not adopted for the present investigations,
analytical continuation schemes for each individual term
of the underlying  KKR-NEGF expression  
also appear viable, after adaptation from the retarded GF case \cite{Eschrig1986}.

\section{Absorption spectra via retarded GF and via greater GF}

\label{sec-abs}

\begin{figure}
\includegraphics[width=8.5cm]{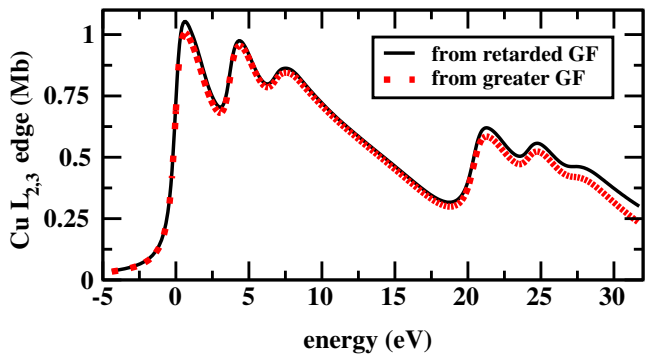}
\caption{\label{fig:XAS-retard-greater}
Calculated XAS spectra for the $L_{2,3}$-edge of pure Cu
based on the standard expression
for $\mu^{\lambda}(\omega)$  shown using the retarded GF 
(full black curve)
and the generalized expression in Eq.~(3) in the main text shown
using the greater GF 
(dashed red curve).
}
\end{figure}

The expression for the 
absorption coefficient 
$\mu^{\lambda}_{\Phi}(\omega)$
given in terms 
of the greater GF in Eq.~(3) in the main text is applicable also 
for the equilibrium case ($\Phi=0$).
This offers the opportunity to check the numerical results based on Eq.~(3)
with results based on the standard expression using the retarded GF
\cite{Ebert1996}.   
Fig.~\ref{fig:XAS-retard-greater} shows a corresponding comparison 
for the $L_{2,3}$-edge of pure Cu, that 
confirms the correctness of our generalized expression
for calculating the absorption coefficient.

\section{Edelstein effect calculations}

\label{sec-edel}

In the linear response regime, the static spin polarization induced by an
external electric field applied to a noncentrosymmetric solid is described by
the so-called Edelstein or inverse spingalvanic effect \cite{Aronov1989,Edelstein1990}.
Here spin-orbit coupling (SOC) and absence of inversion symmetry lead to an induced
spin polarization of the conduction electrons.

Using Kubo's linear response formalism the Edelstein response tensor
$\underline{\boldsymbol{p}}$ that gives the relation between spin polarization
$\boldsymbol{s}$ and electric field $\boldsymbol{E}$, $\boldsymbol{s} =
\underline{\boldsymbol{p}} \boldsymbol{E}$, can be given by an expression analogous 
to the Kubo-Bastin formula for the conductivity tensor \cite{Crepieux2001},
\begin{align}
  p_{\mu\nu}
  =&
    - \frac{\hbar }{4\pi}\int_{-\infty}^{\infty}d\varepsilon
    \frac{df(\varepsilon)}{d\varepsilon}
     \mathrm{Tr} \left[
       \hat{P}_{\mu}(G^{r}-G^{a})  \hat{j}_{\nu} G^{a} -
       \right.     \nonumber  \\
       &
       \vspace*{3em}
       \left.
       \hat{P}_{\mu} G^{r} \hat{j}_{\nu}(G^{r}-G^{a})
     \right] 
   \nonumber  \\
  & 
    + \frac{\hbar }{4\pi}\int_{-\infty}^{\infty}d\varepsilon
    f(\varepsilon)
    \mathrm{Tr}
    \left[
        \hat{P}_{\mu}G^{r}\hat{j}_{\nu}
        \frac{dG^{r}}{d\varepsilon}
        - 
        \right.
         \nonumber  \\
        &
        \vspace*{3em}
        \left.
        \hat{P}_{\mu}\frac{dG^{r}}{d\varepsilon}  \hat{j}_{\nu} G^{r} 
        - \hat{P}_{\mu}G^{a}\hat{j}_{\nu}
        \frac{dG^{a}}{d\varepsilon}
        -\hat{P}_{\mu}\frac{dG^{a}}{d\varepsilon}  \hat{j}_{\nu} G^{a}        
    \right]
    \label{eq:Edelstein-Bastin1}
    \; ,
\end{align}
%
where  $f(E)$ 
is the Fermi-Dirac distribution, 
evaluated here for the same chemical potential $\mu=E_F$ at $T=0$~K
throughout the trilayer system, 
$G^{r}$ ($G^{a}$) is
the retarded (advanced) single-particle Green function at energy $ E$
(arguments have been dropped for the sake of readability),
 $\hat{P}_{\mu}$ is the spin-polarization operator and
  $\hat{j}_{\nu}$ is the current density operator (see below). 
 The presence of $f(E)$
implies that in
the temperature limit $T \rightarrow 0$~K the first term in
Eq.~\eqref{eq:Edelstein-Bastin1} above has to be evaluated only for
the Fermi energy 
$E_F$ (Fermi surface term), 
while the second one
requires an integration over the occupied part of the valence band  (Fermi sea
term),
which has been omitted based on the experience 
with the anomalous and spin Hall
coefficient \cite{KCE15}.
The product of two Green functions in 
Eq.\ \eqref{eq:Edelstein-Bastin1} implies that the response in an atomic cell $n$
is due to the perturbation in all cells $n'$. This leads in a natural way to
the site resolved response function  $p_{\mu\nu}^{nn'}$ mentioned in
the main text.

The current (density) operator $\hat{j}_{\nu}= - |e| c \alpha_\nu
$ in 
Eq.~\eqref{eq:Edelstein-Bastin1} represents the perturbation due to the
electric field component $E_\nu$.  Adopting a fully relativistic formulation to
account coherently for the impact of SOC,  $\hat{j}_{\nu}$ is expressed by the
corresponding velocity operator  $\hat{v}_{\nu}= c \alpha_\nu$, where $c$ is
the speed of light and  $\alpha_\nu$ is one  of the standard $4 \times 4 $
Dirac matrices \cite{Rose1961}. The spin-polarization operator $\hat{P}_\mu$
on the other hand represents the spin-magnetic moment along the $\mu$~axis
induced by the electric field, and is therefore most conveniently expressed
in a relativistic four-component Dirac
notation by:
%
\begin{equation}
\label{EE-operator}
\hat{P}_{\mu} = 
\beta
\Sigma_\mu  = 
\left(
\begin{array}{cc} \sigma_\mu & 0_2 \\ 0_2 & -\sigma_\mu \end{array} \right) ~,
\end{equation}
%
with $\sigma_\mu$ being one of the standard 2~$\times$~2 Pauli matrices \cite{Rose1961}.

Figure 3 in the main text 
shows results obtained 
by adding an imaginary part of  $10^{-5}$~Ry to the
Fermi energy and assuming $T=0$~K.

\section{Evaluating the sum rule}

\label{sec-sumrule}

Making use of the modified sum rules
given in Eq.~(4) in the  main text,
similar considerations
concerning the energy path and integration
as outlined above
also apply to the greater GF,
which is used to compute the XAS/XMCD spectra
in the non-equilibrium situation.
Far above the Fermi level,
results remain unmodified 
with respect to a retarded GF calculation,
obtained under direct inclusion of the layer-to-layer voltage drop 
within each atomic potential. 
This allows to invest more computational effort
for the low-lying contributions from the greater GF. 

 This work deals with layered systems via the tight-binding KKR  
(TB-KKR) formalism, meaning that for spectroscopy application, further 
care is required in choosing a suitably large
(TB-KKR) screening parameter \cite{Zeller1997}.
 Namely, the greater GF needed to calculate the spectra has to be
  evaluated for higher energies than the lesser GF needed to
  calculate the ground state observables [compare Eqs.~(2) and (3) in
    the main text].
The use of the TB-KKR scheme and considering
energies well above the Fermi energy also implies a larger angular momentum cutoff,
as already known from studies based on the retarded GF alone.
Throughout this work, XAS/XMCD calculations were performed by using a TB-KKR screening parameter of $12$ Ry,
and spherical harmonics cut-off at $\ell_{max}=4$.

For the practical use of the sum rule expression, 
one finally has to determine the energy cutoff parameter $E_{\text{cut}}$ 
up to which $10$ $d$ electrons are entirely accounted for.
Similarly as in Ref.~\cite{SME09b,SMS+11}, we find this 
 energy by requiring that the density
of the d states accounts exactly for 10 electrons
 when integrated up to $E_{\text{cut}}$.
For transition metal systems as considered in Ref.~\cite{Kukreja2015},
we are typically in a situation 
where this IDOS value from the sum of lesser and greater GF
lies fairly above the $\left[ E_F - \frac{\Phi}{2}, E_F + \frac{\Phi}{2} \right]$
energy window, 
in which the NEGF self-energy is operative.
Noting the complementarity 
in the definition of lesser and greater Green functions
\cite{Achilles2013,Ogura2016},
we can exploit the DOS relationship:
$n^{<}(E) + n^{>}(E) = n^{r}(E)$,
where all quantities are computed 
under the same assumed voltage drop.
It is then possible to estimate the upper limit 
for absorption spectra integrations 
$E_{\text{cut}}$
in terms of the IDOS from the right hand side term alone.

\bibliography{NEGF-XAS-complete}

\end{document}